\documentclass[twocolumn,prl,aps,showpacs,superscriptaddress]{revtex4}
\usepackage{amsmath,amssymb,eucal,graphicx}
\begin{document}

\title{Rank Statistics in Biological Evolution}
\author{E.~Ben-Naim}
\affiliation{Theoretical Division and Center for Nonlinear
Studies, Los Alamos National Laboratory, Los Alamos, New Mexico
87545, USA}
\author{P.~L.~Krapivsky}
\affiliation{Department of Physics and Center for Molecular Cybernetics,
Boston University, Boston, Massachusetts 02215, USA}

\begin{abstract}
  We present a statistical analysis of biological evolution processes.
  Specifically, we study the stochastic replication-mutation-death
  model where the population of a species may grow or shrink by birth
  or death, respectively, and additionally, mutations lead to the
  creation of new species.  We rank the various species by the
  chronological order by which they originate. The average population 
  $N_k$ of the $k^{\rm th}$ species decays algebraically with rank,
  $N_k\sim M^{\mu}k^{-\mu}$, where $M$ is the average total
  population. The characteristic exponent
  $\mu=(\alpha-\gamma)/(\alpha+\beta-\gamma)$ depends on $\alpha$,
  $\beta$, and $\gamma$, the replication, mutation, and death
  rates. Furthermore, the average population $P_k$ of all descendants
  of the $k^{\rm th}$ species has a universal algebraic behavior,
  $P_k\sim M\,k^{-1}$.
\end{abstract}
\maketitle

Darwin's seminal ideas provide a qualitative basis for understanding
biological evolution. Yet, quantitative characterization of evolution
involves formidable challenges. For example, there is a large
uncertainty in the total number of species existing on Earth, with
estimates ranging from 2 to 100 million \cite{life}.  Furthermore,
only a small fraction of all of the extinct species is presently known
\cite{primate}. Biological evolution remains one of the deepest and
most elusive problems in science.

Fossil records \cite{gould,sep} and molecular sequences data
\cite{msw,dekm} provide quantitative clues to understanding
evolution. They are widely used to chronicle evolution processes using
``trees of life'' \cite{tol}, evolutionary trees that document how
species, viruses, humans, etc., are related \cite{page,fel}.
Reconstruction of evolutionary trees from existing sequences is a
challenging problem \cite{msw,dekm} because only partial information
is available, and because one has to determine the most plausible
evolution history from an exponentially large number of scenarios. But
the most significant difficulty is that the evolution laws themselves
are unknown. In practice, tree reconstruction methods rely heavily on
simplified evolution models: branching processes \cite{yule,simon,teh}
incorporating elementary processes such as replication, mutation,
death, etc.

We study the standard Replication-Mutation-Death (RMD) process that
has been widely used to model speciation \cite{primate}, population
genetics \cite{ewens}, and genome evolution
\cite{acl,wl,gll,bl,rh,hv,kwk,jlg,ds,mal,snel}. The RMD process
incorporates the minimal mechanisms for evolution: a family of species
grows and shrinks by natural birth and death,
respectively. Additionally, mutations create new families of species.

We focus on the rank, namely, the chronological order by which the
species are created. We study the total population size and the
total descendant population size of a species of a given rank. Our
main result is that both these quantities decay algebraically with the
rank. While the scaling law characterizing the population size depends
on the details of the evolution process, the scaling law
characterizing the descendant population size is universal. We
conclude that the chronological rank provides a useful characteristic
of evolution.

The Replication-Mutation-Death (RMD) process is defined as follows. At
any given time there are multiple families of distinct species and
each species family has a certain size. Evolution proceeds as
organisms (i) replicate: give birth to an identical child, (ii)
mutate: give birth to a mutant, or (iii) die: are removed due to
death. Replication events increase the size of the corresponding
species family by one, and similarly, death events reduce the family
size by one. In this minimal model, the population is asexual and
therefore, replication mimics asexual reproduction. The key feature of
this model is that each mutation event creates a distinct new
species. All three processes are completely random and independent of
each other. Initially, there is only one organism, and hence, only one
species. We term the first species, the grand ancestor.

Let $\alpha$, $\beta$, and $\gamma$ be the rates of replication,
mutation, and death, respectively. The total growth rate must of
course be positive, $\alpha+\beta-\gamma>0$. Moreover, we restrict our
attention to the biologically relevant case where the replication rate
exceeds the death rate, $\alpha>\gamma$. The most elementary
characteristic is the total population size. The average total
population size $M$ grows according to $dM/dt=(\alpha+\beta-\gamma)
M$, and given $M(0)=1$, the total population increases exponentially
with time
\begin{equation}
\label{M}
M(t)=e^{(\alpha+\beta-\gamma)t}\,.
\end{equation}
Throughout this study, averaging is taken with respect to infinitely
many independent realizations of the stochastic process.

The RMD evolution process is illustrated in Fig.~1. One natural
characteristic is the ``distance'', the number of mutation events that
separate a descendant and its ancestor.  In Fig.~1, the distance
between species 3 and species 1 equals 2 and the distance between
species 4 and species 1 is 1.  Let $G_n$ be the total population with
distance $n$ from the grand ancestor.  This quantity evolves according
to
\begin{equation}
\label{Gn-eq}
\frac{dG_n}{dt}=(\alpha-\gamma)\,G_n+\beta\,G_{n-1}\,.
\end{equation}
The initial condition is $G_n(0)=\delta_{n,0}$. Equation (\ref{Gn-eq})
reflects that the distance is augmented by one with each mutation. It
is convenient to normalize $G_n$ by the total population size,
$G_n=M\,g_n$.  The quantity $g_n$ satisfies $dg_n/d(\beta
t)=g_{n-1}-g_n$ with $g_n(0)=\delta_{n,0}$. Solving this equation, the
probability distribution $g_n$ is Poissonian, $g_n(t)=(\beta
t)^n\,e^{-\beta t}/n!$, and therefore
\begin{equation}
\label{Gn}
G_n(t)=\frac{(\beta t)^n}{n!}e^{(\alpha-\gamma)t}\,.
\end{equation}
The Poissonian distribution of distance reflects the completely random
nature of the evolution process. 

\begin{figure}[t]
\includegraphics*[width=0.3\textwidth]{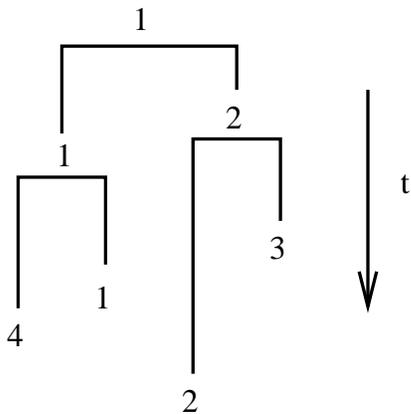}
\caption{Illustration of the replication-mutation-death process with a
total of 4 different families of species and a total population of 7
(here, there are no death events).  The rank of each species is
indicated.}
\end{figure}

Let $S(t)$ be the average number of mutation events until time
$t$. This quantity follows easily from the average population size,
\begin{equation}
\label{S-eq}
\frac{dS}{dt}=\beta M.
\end{equation}
For the initial condition $S(0)=0$, the average number of mutation
events, or equivalently, the average number of distinct species
generated is
\begin{equation}
\label{S}
S(t)=\frac{\beta}{\alpha+\beta-\gamma}
\left[e^{(\alpha+\beta-\gamma)t}-1\right].
\end{equation}
Therefore, the total number of distinct species originated throughout
the evolution quickly becomes of the same order as the total
population size: $S/M\to \beta/(\alpha+\beta-\gamma)$ for $t\gg
1/(\alpha+\beta-\gamma)$. We note that the total number of existing
species is generally smaller than the total number of species created
because some species become extinct.

The chronological order by which the species are created is a useful
way to characterize the evolution process (Fig.~1). The grand ancestor
has the index $k=1$, the second species (generated by the first
mutation event) has the index $k=2$, etc. We term this chronological
index, the rank. With this definition, the rank of an ancestor is
always smaller or equal than the rank of any of its descendants. We
note that different definitions of rank are used to characterize data
storage trees in computer science \cite{dek} and river networks in
geophysics \cite{reh}.  
Statistics of the maximal rank are characterized by $f_k$, the
probability that the total number of distinct species originated up to 
time $t$ equals $k$.  The distribution $f_k$ satisfies the evolution
equation
\begin{equation} 
\label{Fk-eq}
\frac{df_k}{dt}=\beta\, M\,\left(f_{k-1}-f_k\right).
\end{equation}
The initial condition is $f_k(0)=\delta_{k,1}$ and the boundary
condition is $f_{0}(t)=0$.  This equation reflects that every
mutation event creates a new species. We stress that this evolution
equation is not exact: it approximates the total population size, a
fluctuating quantity, by its average value \cite{comment}. In other
words, it assumes that the total population size and the total number
of distinct species are not correlated. Nevertheless, as shown below,
this is an excellent approximation. Using Eq.~(\ref{S-eq}), the overall
multiplicative factor $\beta M$ in Eq.~(\ref{Fk-eq}) is eliminated
by redefining the time variable, \hbox{$df_k/dS=f_{k-1}-f_k$}. The
resulting probability distribution is Poissonian,
\begin{equation}
\label{Fk}
f_k(t)=\frac{S^{k-1}}{(k-1)!}e^{-S}\,,
\end{equation}
again reflecting the random nature of the evolution process. By
construction, the first two moments are exact, $\sum_k f_k=1$ and
$\sum_k k f_k=1+S$.

The distribution $f_k$ enables us to determine various statistical
properties of the rank. A natural question is: what is the population
of a species of a given rank?  For example, in figure 1, the
population of the first species equals 3 and the population of the
second species equals 2.  Consider $N_k$, the average population size
of the $k^{\rm th}$ species. It evolves according to
\begin{equation}
\label{Nk-eq}
\frac{dN_k}{dt}=(\alpha-\gamma)\, N_k+\beta\,M\, f_{k-1},
\end{equation}
with the initial condition $N_k(0)=\delta_{k,1}$. The first term on
the right-hand side accounts for replication and death while the
second term describes mutations.  As in Eq.~(\ref{Fk-eq}), this
equation is approximate: in writing the second term, we assumed that
the total population size and the total number of species are not
correlated. Writing $N_k=e^{(\alpha-\gamma)t}n_k$ transforms
Eq.~(\ref{Nk-eq}) into $dn_k/dt=\beta\,e^{\beta t}\,f_{k-1}$.  For the
grand ancestor $n_1=1$, and for $k\geq 2$
\begin{equation}
\label{Nk}
n_k(t)=\beta\,\int_0^t d\tau\, f_{k-1}(\tau)\,e^{\beta \tau}\,.
\end{equation}

To obtain the large-rank behavior ($k\gg 1$) for large populations
($t\gg 1$) we use asymptotic analysis
\begin{eqnarray*}
n_k&=&\beta \int_0^t d\tau\, f_{k-1}(\tau)\,e^{\beta\tau}\\
   &=&\frac{1}{(k-2)!}\int_0^{S} dx\, \left[1+\frac{x}{1-\mu}\right]^{-\mu}
x^{k-2} e^{-x}\\
   &\to&\frac{1}{(k-2)!}\int_0^{\infty} dx\,
\left[1+\frac{x}{1-\mu}\right]^{-\mu}
x^{k-2} e^{-x}\\
   &\sim& k^{-\mu},
\end{eqnarray*}
with $\mu=(\alpha-\gamma)/(\alpha+\beta-\gamma)$.  The second line was
obtained by substituting (\ref{S}) and (\ref{Fk}) into (\ref{Nk}) and
the third line was obtained by replacing the upper integration limit
by infinity because $S$ diverges rapidly. The integral is estimated by
the steepest descent method using the fact that the integrand is
maximal for $x\approx k$.  Therefore, the average total population of
a species decays algebraically with its rank (Fig.~2)
\begin{equation}
\label{Nk-large}
N_k\sim M^\mu\,k^{-\mu}, 
\end{equation}
for $k\gg 1$.  The characteristic exponent 
\begin{equation}
\label{mu}
\mu=\frac{\alpha-\gamma}{\alpha+\beta-\gamma},
\end{equation}
is positive, $\mu>0$, since the replication rate exceeds the death
rate.  For the special case of no-death, $\gamma=0$, it is possible to
derive the power-law behavior (\ref{Nk-large})-(\ref{mu}) exactly by
using the total population rather than time to characterize the
evolution (this leads to exact difference evolution equations rather
than the differential equation (\ref{Nk-eq}) \cite{bk}). When the
replication rate approaches the death rate, the population size
becomes rank-independent.  The characteristic exponent
\hbox{$0<\mu\leq 1$} varies continuously with the three rates.

\begin{figure}[t]
\includegraphics*[width=0.4\textwidth]{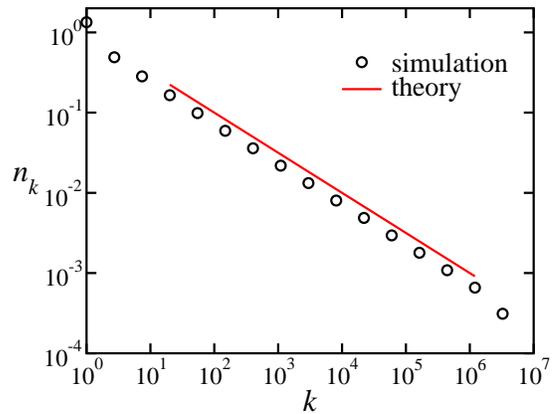}
\caption{The normalized size distribution $n_k=M^{-1/2}N_k$ versus
$k$, obtained from $10^3$ independent Monte Carlo realization for a
total population of size $10^7$. The rates are $\alpha=\beta$ and
$\gamma=0$. Also shown for comparison is the theoretical prediction
$n_k\sim k^{-1/2}$.}
\end{figure}

To test the theoretical predictions, we performed Monte Carlo
simulations.  We focus on the no-death case ($\gamma=0$) because it
can be simulated efficiently, thereby allowing us to generate
populations of size up to $10^7$. The simulation is straightforward:
an organism is picked at random. Then, with probability
$\alpha/(\alpha+\beta)$ an identical organism is created
(replication), while with probability $\beta/(\alpha+\beta)$, a new
species is created (mutation).  The simulation results are in
excellent agreement with the theoretical results. For example, for the
case $\alpha=\beta$, the exponent agrees with the theoretical
prediction $\mu=1/2$ to within $0.1\%$.  We also simulated the case
where the death rate is finite and the agreement between the
simulations and the theoretical predictions concerning the exponent
$\mu$ was equally strong. We conclude that the numerical simulations
indicate that even though the rate equations (\ref{Fk-eq}) and
(\ref{Nk-eq}) are approximate, they yield asymptotically exact results
for the large-rank behavior. In particular, the predicted exponent
$\mu$ appears to be exact.

We also noticed that in a single realization, $N_k$ deviates
significantly from the theoretical prediction, indicating that there
are strong sample-to-sample fluctuations. This is not surprising
because the populations are growing exponentially. To extract
meaningful results from a single realization, it is necessary to study
statistics of the variable $z=\ln k$. Using this
exponentially-distributed variable, the large fluctuations are
suppressed, and the power-law behavior (\ref{Nk-large}) is clear.

A closely-related quantity is the total descendant population of a
species, that is, the total population of all species that emanated
from a given species. For example, in figure 1, the total descendant
population of the first species is 7 and that of the second species is
3. Fixing the $k^{\rm th}$ species as a common ancestor, we denote by
$P_k$ the average size of its descendant population.  This quantity
satisfies the initial condition $P_k(0)=\delta_{k,1}$ and it evolves
according to
\begin{equation}
\label{Pk-eq}
\frac{dP_k}{dt}=(\alpha+\beta-\gamma)\,P_k+\beta M f_{k-1}\,.
\end{equation}
Since a descendant of a descendant is also a descendant, the growth
rate now accounts for mutation too, in contrast with
Eq.~(\ref{Nk-eq}).  The transformation $P_k=M\,p_k$ recasts
Eq.~(\ref{Pk-eq}) into $dp_k/dt=\beta f_{k-1}$. Since every species is 
the descendant of the grand ancestor, $p_k=1$. Otherwise, for $k\geq
2$ one has
\begin{equation}
\label{Pk}
p_k(t)=\beta\,\int_0^t d\tau\, f_{k-1}(\tau).
\end{equation}
The quantity $p_k$ is of course the probability that a randomly
selected organism is a descendant of the $k^{\rm th}$ species.
Repeating the asymptotic analysis that has been applied to
Eq.~(\ref{Nk-eq}), we find that $p_k(t)$ reaches an asymptotic value,
$p_k(t)\to p_k(\infty)$ as $t\to\infty$, with 
\begin{equation*}
p_k(\infty)=\frac{1}{(k-2)!}\int_0^{\infty} dx\,
\left[1+\frac{x}{1-\mu}\right]^{-1} x^{k-2} e^{-x}\,.
\end{equation*}
In the large-rank limit we have $p_k\sim k^{-1}$, and therefore
\begin{equation}
\label{Pk-large}
P_k\sim M\,k^{-1}
\end{equation}
(Fig.~3). Thus, the descendant size distribution obeys a universal law
as the characteristic exponent is independent of the various rates.

\begin{figure}[t]
\includegraphics*[width=0.4\textwidth]{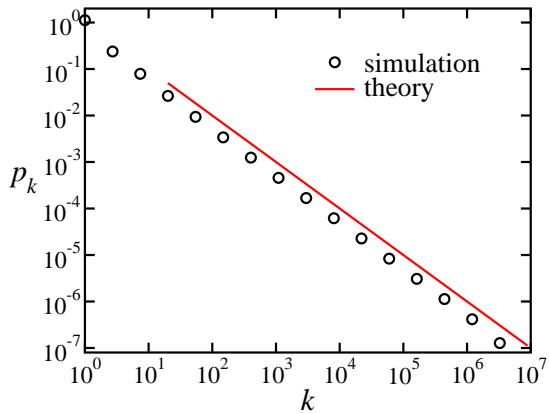}
\caption{The descendant size distribution. Plotted is $p_k=M^{-1}P_k$
  versus $k$, obtained from $10^3$ independent Monte Carlo realization
  for a population of size $M=10^7$. The death rate is zero
  ($\gamma=0$), while replication and mutation rates are equal
  ($\alpha=\beta$). Also shown for comparison is the theoretical
  prediction $p_k\sim k^{-1}$.}
\label{nk}
\end{figure}

In summary, we obtained scaling laws that characterize how the
population of a species and its descendant population depend on the
rank. The average population of a species decays algebraically with
the rank $k$, $N_k\sim k^{-\mu}$, with the characteristic exponent
$\mu$ dependent on the rates of replication, mutation, and death. The
average descendant population of a species decays in a universal
fashion with the rank $k$, $P_k\sim k^{-1}$. Our main conclusion is
that the chronological order of origination, or the rank, provides a
useful characterization of biological evolution processes.

Several aspects of this model should be investigated further.
Obtaining the exact behavior by properly accounting for the coupling
between the total population size and the total number of species is a
challenging problem. We focused on the average behavior, but
statistical fluctuations must exhibit interesting behavior because the
population is exponentially-growing.

Power-law distributions with different rate-dependent exponents were
found for another quantity, the family size distribution, for the very
same replication-mutation-death processes
\cite{ds,rh,hv,kwk,jlg}. In general, the various rates can not be
measured directly. Since the characteristic exponents yield a
constraint for the rates, such scaling-laws are useful since they
reduce the number of unknown parameters in the problem.

\acknowledgments
We acknowledge US DOE grant W-7405-ENG-36 for support of this
work.


\begin{thebibliography}{99}
\bibitem{comment} An exact description must involve the joint population
    size-mutation number probability distribution.
\end{thebibliography}
\end{document}